\documentclass[graybox]{svmult}

\usepackage{mathptmx}       
\usepackage{helvet}         
\usepackage{courier}        
\usepackage{type1cm}        
\usepackage{makeidx}         
\usepackage{graphicx}        
\usepackage{multicol}        
\usepackage[bottom]{footmisc}

\makeindex             

\begin{document}

\def \mum{$\mu$m}
\newcommand \kms{km~$\rm{s}^{-1}$}
\newcommand \cc{$\rm{cm}^{-3}$}
\newcommand \lam{$\lambda$}
\newcommand \lsol{L$_{\odot}$}
\newcommand \msol{M$_{\odot}$}
\newcommand \rsol{R$_{\odot}$}
\newcommand \mdot{M$_{\odot}$ yr$^{-1}$}
\newcommand \fdens{erg s$^{-1}$ cm$^{-2}$ arcsec$^{-2}$}
\newcommand \flux{erg s$^{-1}$ cm$^{-2}$}
\newcommand \lum{erg s$^{-1}$}
\newfont{\rten}{cmr10} 
\def \arcdeg{\hbox{$^\circ$}}
\def \arcmin{\hbox{$^\prime$}}
\def \arcsec{\hbox{$^{\prime\prime}$}}

\title*{Star Formation in the Milky Way. The Infrared View.}
\author{Alberto Noriega-Crespo}
\institute{A. Noriega-Crespo \at IPAC, California Institute of Technology,\email{alberto@ipac.caltech.edu}}
%
%
\maketitle

\abstract*{I present a brief review of the some of the most 
recent and active topics of star formation process in the Milky Way 
using mid and far infrared observations, and motivated by 
the research being carried out by our science group using the data gathered by the Spitzer 
and Herschel space telescopes. These topics include bringing together the scaling relationships
found in extragalactic systems with that of the local nearby molecular clouds, the 
synthetic modeling of the Milky Way and estimates of its star formation rate.}

\abstract{I present a brief review of the some of the most 
recent and active topics of star formation process in the Milky Way 
using mid and far infrared observations, and motivated by 
the research being carried out by our science group using the data gathered by the Spitzer 
and Herschel space telescopes. These topics include bringing together the scaling relationships
found in extragalactic systems with that of the local nearby molecular clouds, the 
synthetic modeling of the Milky Way and estimates of its star formation rate.}

\section{Introduction}
\label{sec:1}

In December 2007 we, Kartik Seth and I, organized a meeting in Pasadena called
``The Evolving ISM in the  Milky Way and Nearby Galaxies'' 
(Seth, Noriega-Crespo, Ingalls \& Paladini [26]); it was the
fourth conference under the auspices of the Spitzer Science Center. The main
goal of that meeting was to bring together two communities using the same 
tools, methodology and looking essentially at the same issues, with the only 
differences being on the spatial scales and samples of astrophysical objects 
that they were studying. 
Two of the leading participants were Dr. Neal Evans and Dr. Robert Kennicutt, 
both Principal Investigators of two very successful  Spitzer Legacy Surveys,
 ``c2d'' ('Core to Disks')  and ``SINGS'' ('The Spitzer Infrared Nearby 
Galaxies Survey'), respectively. Both interested on the star formation process,
one locally (Evans) and one on nearby galaxies (Kennicutt). Therefore, it was
not a surprise that the same day I gave this presentation at Sant Cugat in the
workshop on {\it Cosmic-ray induced phenonmenology in star-forming 
environments}, that Kennicutt and Evans submitted to astroph [14]
 a review on ``Star Formation in the Milky Way and Nearby Galaxies'' to appear
in the Annual Reviews of Astronomy and Astrophysics during the Fall of 2012 [14].
And although their review is certainly more ambitious and complete than this summary, 
I was quite pleased to see that we identified some of the same main issues 
and progress on the subject. 

Understanding the star formation process locally or in extragalactic systems 
is a very active research area, and where infrared observations have played a 
major role in bringing together a wholistic view. Measuring the star formation
 rate (SFR) of the Milky Way (MW) or other galaxies is like ``taking their pulse'' 
(Chomiuk \& Povich [7]), since the transformation of molecular gas into
stars, plus the energetics and evolution of the massive stars, sets some of 
the main characteristics of what we can observe, e.g. the interstellar medium
chemical composition, the overall gas mass and the bolometric flux densities
at different wavelengths. From the point of view of the Cosmic Ray community,
the interest on star formation and what has been learned at infrared 
wavelengths is quite clear, since massive stars, their fast evolution and 
transformation into supernovae, is one of the main sources of cosmic ray 
acceleration.

This contribution will follow a similar path as that of the original oral
presentation, starting with the star formation rate esitmates from nearby 
clouds (\S1), connecting these estimates with those obtained from the 
extrapolation of the SFR extragalactic indicators (\S2), using then the concept
of a high density molecular gas threshold to connect local and extragalactic
measurements (\S3), follow by the analysis of synthetic modeling of the SFR
in the Milky Way (\S4) and finally looking at what the latest far-infrared
measurements of the Galactic Plane from Herschel are telling us on the
SFR (\S5).

Stars need to form from gas, and although this may seem too obvious today,
 it was Schmidt  [27] nearly fifty years ago who suggested it, 
 by looking at the distribution of the Population I
stars in the Milky Way, that stars form from HI gas and that the rate of star
formation (SFR) was proportional to the square of the volumetric gas density,
i.e. $SFR \propto \rho^2$. Stars actually form in molecular clouds, where
H$_2$ is the dominant specie, and this is quite relevant because there is not in our
Galaxy or other extragalactic systems a one-to-one correlation between the spatial 
distribution of HI and that of H$_2$, in other words Schimdt did not get it quite right.
Nevertheless, this prescription, allowed him to study fundamental properties of the Galaxy,
like its luminosity function, the spatial distribution of stars and its chemical 
evolution [27].
Another key characteristic of the star formation process, it that the process 
itself seems to be different for low mass stars (0.1 - 8 \msol) and high mass stars
($\geq$ 8\msol). And although there is a consensus on the main phases of the accretion process
that leads to the formation low mass stars, this is not the case for high mass protostellar 
objects (see e.g. [1]).

%
\begin{figure}[b]

\includegraphics[scale=.65]{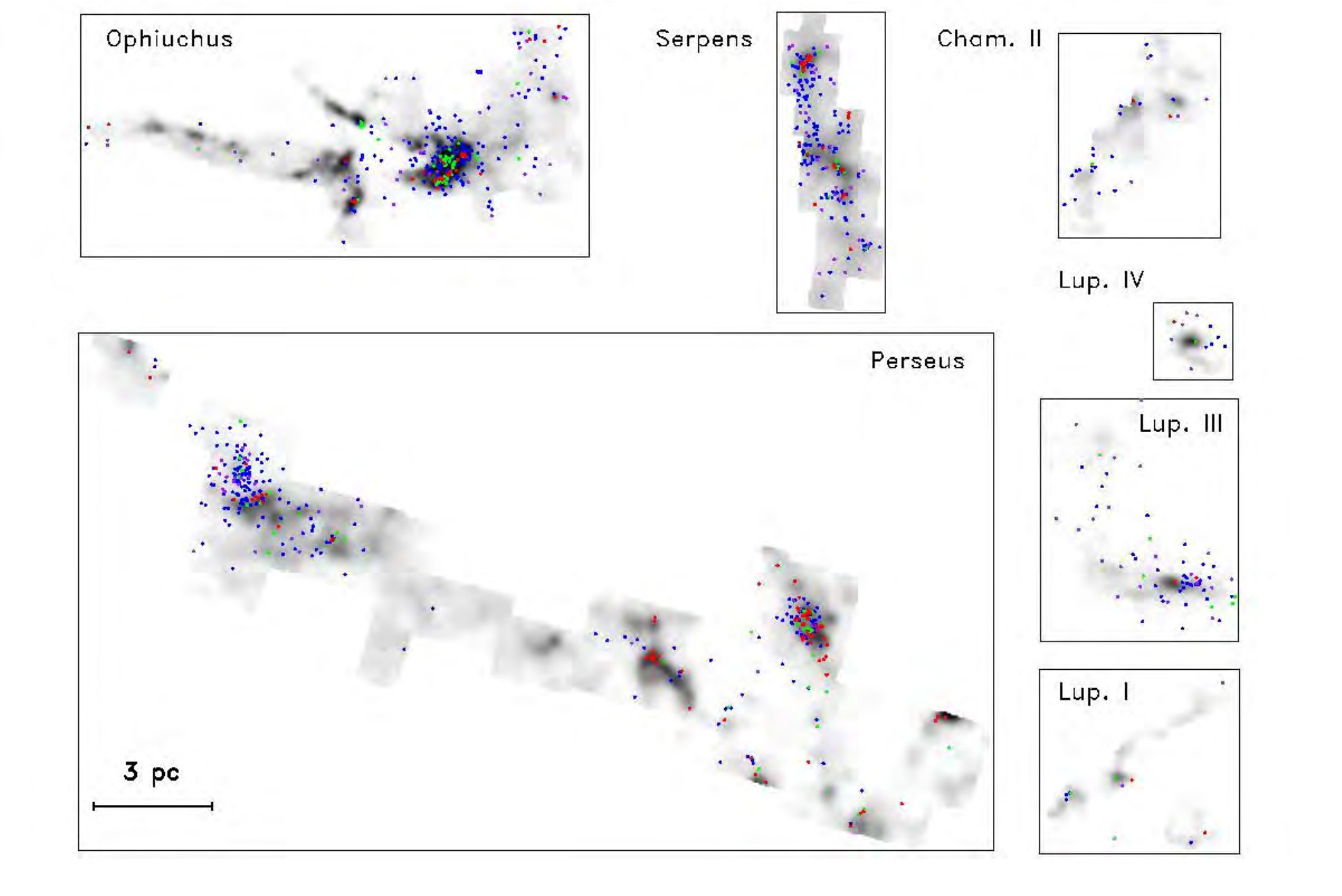}
%
%
\caption{The seven nearby star forming molecular clouds observed by the c2d team to
determine their star formation rate (from Evans et al. [10]), defined
by the extend of their extinction (grayscale) maps from A$_V$  = 1 to 25 mag. 
The colored points correspond to the different classes of low mass young stellar objects,
I (red), II (blue) and III (purple), within them.}
\label{fig:1}       
\end{figure}

\section{From Local to Global Star Formation Rates and Efficiencies}
\label{sec:2}

 A handful of studies have been published by the Galactic astronomers on the
star formation rates (SFR) and efficiencies over the past three years that 
have tried to connect what is measured in the local molecular clouds 
with the results and relationships obtained by the extragalactic groups. 
The extragalactic scaling relations are of course based on the Scmidtt-Kennicutt 
"law" that describes the star formation rate (per unit area) as a function of 
the gas mass of the system (gas surface density). Indeed, it is our understanding 
of what 'kind' of gas mass is truly involved in the star formation process, that has evolved
since the time that Schmidt postulated a relation between the SFR as a power 
of the volumetric gas density of the neutral Hydrogen (HI) (see e.g. [12]).

The star formation process in nearby clouds, within 500pc or so from the Solar
neighborhood, can be studied in great detail. Not only one can count the 
exact number of young stars that are formed in each cloud, but also one can
determine their evolutionary stage, and therefore to have a complete picture
of the process. One of the well known disadvantages of using the 'local' 
clouds, is that except for Orion, all of them are tracing the low mass 
star formation, with a median value of $\sim 0.5$\msol over a time scale
of $\sim 2.1\times 10^{6}$ yr [16]. Massive star
formation does take place in the Milky Way, but the larger distances to these
star forming regions, at least of couple of kiloparsecs away, skews our view 
of the star formation process towards the highest mass young stellar objects
(see $\S4$, Fig. 7), and therefore, our view still is incomplete.

%
\begin{figure}[t]
\sidecaption[t]
\includegraphics[scale=0.30]{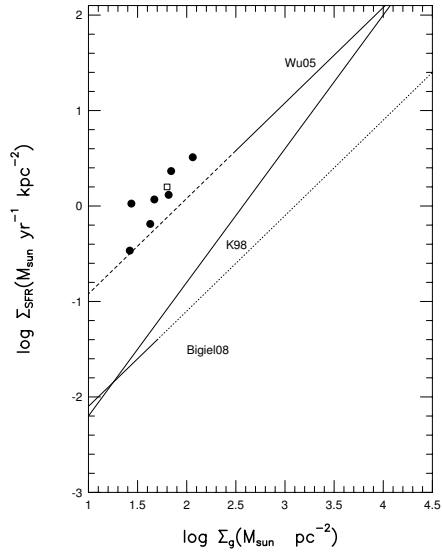}
\caption{The specific star formation rate as a function of specific gas density
for the c2d local star forming clouds. The open square is the average of the seven
clouds (filled circles). The different lines show the SFR relationships derived by
Kennicutt [13], Bigiel et al. [2], using CO as the dense gas tracer, and 
Wu et al. [29], using HCN for tracing the dense gas (from Evans et al. [10])}
\label{fig:2}       
\end{figure}

One of the leading groups studying the star formation process taking place 
nearby is that of the Spitzer Legacy from Cores to Disk (c2d) 
(Evans et al. 10]). This group not only have used the infrared detectors on
board the Spitzer Space Telescope, but has supported their data with
both optical, near-IR and sub-millimetric observations.  The c2d group has
analyzed the properties of five cold clouds (Cha II, Lupus, Perseus, Serpens 
and Ophiuchus; see Fig. 1) and found current star formation rates efficiencies 
ranging from 3\% to 6\%, with a accumulated rate of the five clouds of
$2.6\times 10^{-4}$\msol yr$^{-1}$. Even in this small sample of clouds, the
variation in efficiency strongly suggests that the SF process changes from
cloud to cloud. Furthermore, they found a star formation
rate per unit area (or specific SFR) at least ten times larger than that 
predicted by extrapolating the extragalactic Schmidt-Kennicutt relation to 
their low mass range (see Fig. 2.), 
and they have interpreted this discrepancy as due to the fact that 
at large scales, like those in extragalactic systems, one includes both high
and low density gas in the mass estimates (or gas surface density). This last
conclusion was partially motivated by a previous study by Wu, Evans et al. [29],
where a high gas density tracer, HCN J=1-0 (88.63 GHz), was used to measure the
high density gas, rather than the  standard CO J=1-0, to estimate the gas mass 
of Galactic dense cores, nearby spiral galaxies and farther away starburst 
galaxies. By using a high density gas tracer, they found a tight correlation
over 7-8 orders of magnitude, between the IR luminosity, a direct tracer of
the star formation activity, and the dense material, i.e. a Schmidt-Kennicutt
type of relationship that connected the local SF activity in the MW with 
that of extragalactic systems, both normal and highly active.

\begin{figure}
\sidecaption[t]
\includegraphics[scale=0.5]{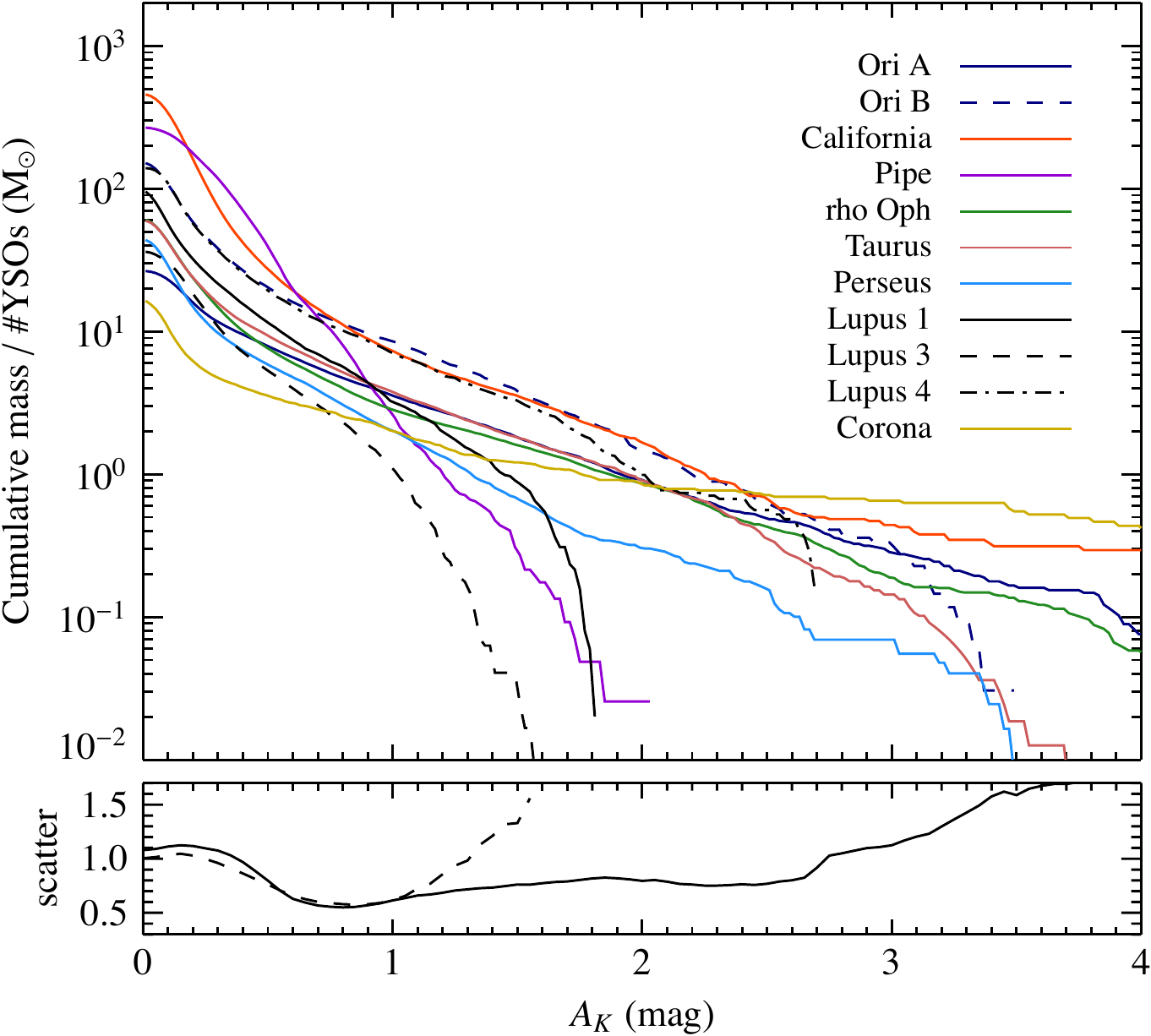}
%
%
\caption{The cumulative mass profile of the clouds in Lada. et al. [16] sample
normalized to their corresponding number of YSOs as a function of infrared extinction.
The eleven clouds reach a minimum at nearly the same magnitude A$_K$ = 0.8 - 0.9mag,
suggesting that there is a threshold in extinction at which the mass contained
is the one most directly involved in the star formation of the cloud.}
\label{fig:3}       
\end{figure}
   
 \section{A High Density Molecular Gas Threshold for Star Formation}
\label{sec:3}

 The Spitzer Legacy surveys were meant to provide a wealth of high quality IR
data that could be used by a wider community to explore topics beyond those
that were originally intended by the project. In the true spirit of taking
advantage of the Legacy data, Lada, Lombardi \& Alves [16] added 
the c2d clouds to their sample of star forming regions for extinction studies in 
the near infrared (NIR). NIR extinction has been successfully used to study the 
total mass of nearby clouds, as well as their structure. Some of the clouds in
the extinction studies are relatively massive, e.g. the Pipe and California
nebulae, but with very little star formation. The sample of extinction studied
clouds doubled  (11 clouds) that of the c2d group, 
and included some of the clouds forming massive
stars like Orion A and B. The extinction method also allows a normalization
of all the clouds to a given extinction threshold, and this idea is quite
powerful since brings the studied molecular clouds on the same mass scale
(see Fig. 3). Lada, Lomabardi and Alves  [16] found that when comparing the star 
formation inventories for these clouds with their extinction masses at a given threshold
in the NIR K band (at $\sim 2.2$\mum) of A$_K$ = 0.8mag, there was a linear
relationship between their SFR and the M$_{0.8}$ mass of the clouds, of the
form SFR(\msol~$yr^{-1}) = 4.6\pm 2.6 \times 10^{-8}$ M$_{0.8}$ (see Fig. 4).

\begin{figure}
\sidecaption[t]
\includegraphics[scale=.50]{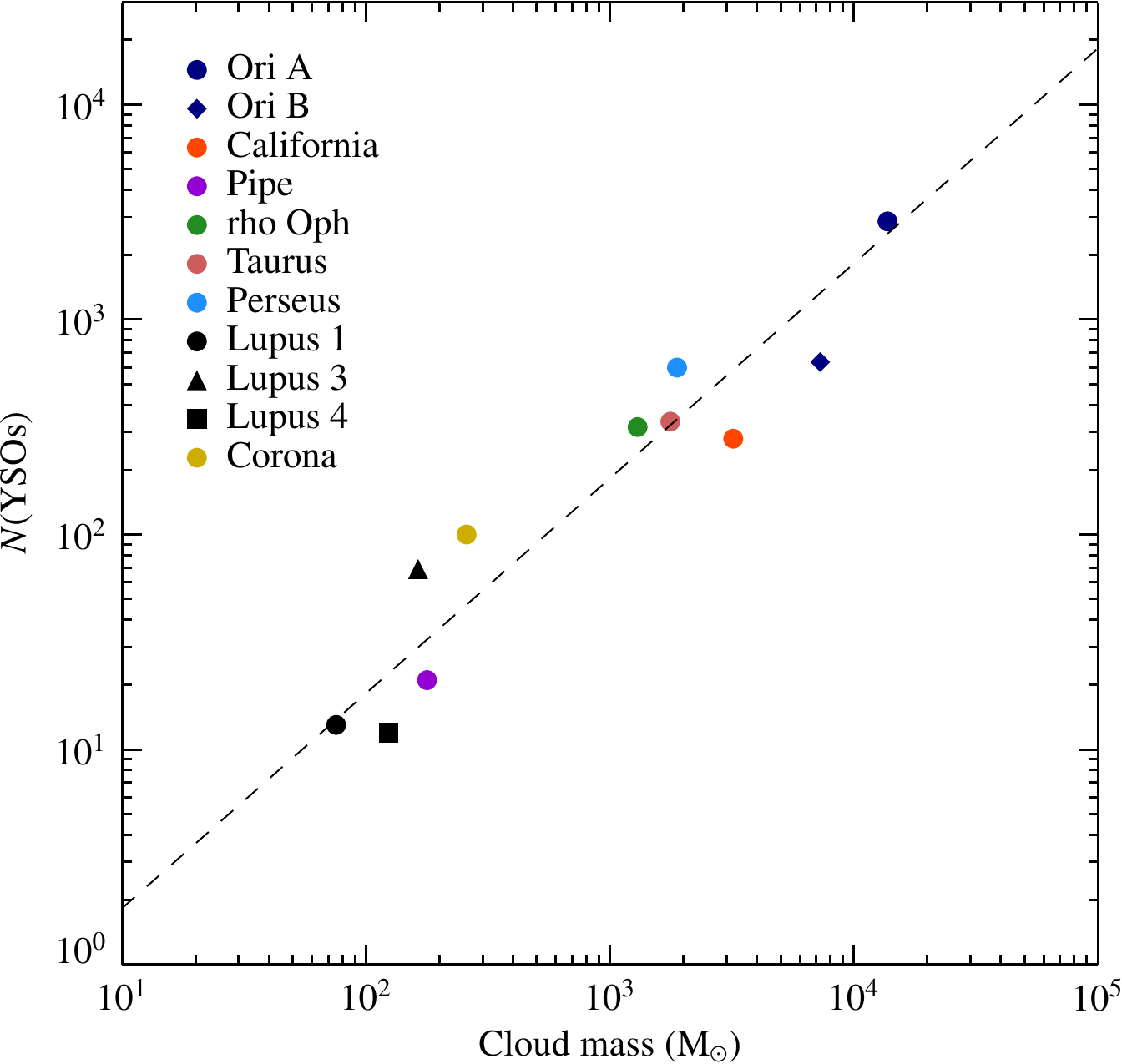}
%
%
\caption{The number of YSOs as a function of the mass of the cloud
set by the A$_K$ = 0.8mag threshold (from Lada, Lombardi \& Alves [16]),
i.e. the mass of the dense gas directly involved with the star formation
process.}
\label{fig:4}       
\end{figure}

 This A$_K$ = 0.8mag threshold corresponds to a gas surface density of
$\Sigma$ = 116\msol pc$^{-2}$, and can be interpreted as a volumetric gas density
threshold in molecular Hydrogen of n(H$_2$) = 10$^{4}$\cc. A similar threshold
was determine by c2d group [11] using the local clouds, 
with $\Sigma$ = 129$\pm$14\msol pc$^{-2}$, where the Galatic SFR versus
the gas mass of the parent cloud becomes linear and not very different
than that of the extragalactic indicators (see Fig. 5). The nice physical
interpretation of these results is that star formation takes place in the densest
regions, above a given threshold, and when this is taking into account the
SFR corresponds to that specific mass of dense gas. Indeed, Lada, Lombardi \&
Alves  [16] suggested that if this is the case, then one should be able
to place in the same relationship, SFR vs dense mass, Galactic Cores and
extragalactic objects, and they explored this possibility in another study [17].

\begin{figure}[t]
\sidecaption[t]
\includegraphics[scale=.35]{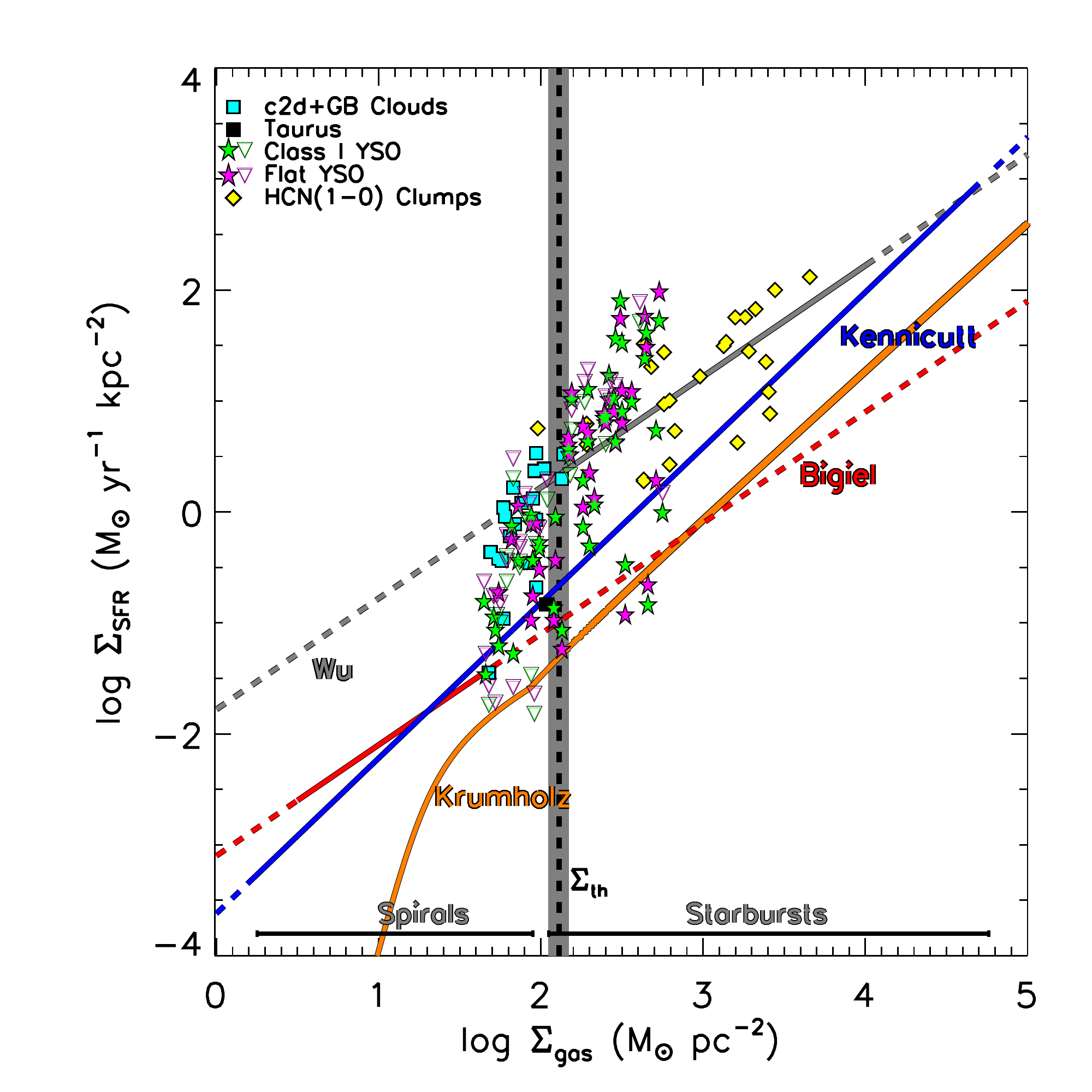}
%
%
\caption{The local Galactic star forming clouds do behave linearly
in the specific star formation rate versus specific gas cloud density diagram,
above a certain threshold ($\Sigma_{th}$), and not very different than
the relationships found for the extragalactic indicators (see 
Heiderman et al. [11] for details).
}
\label{fig:5}       
\end{figure}

Figure 6 shows the SFR vs. cloud molecular mass for the local galactic clouds (circles),
normal galaxies (pentagons), luminous (squares) and ultraluminous (inverted triangles) 
active star formation red galaxies (LIRGS \& ULIRGS) plus high redshift BzK galaxies (triangles).
The open symbols are for measurements of the gas mass based on CO observations ("mean"
molecular masses), while the solid symbols are those based on a dense gas
tracer (e.g. HCN) or extinction. The broken parallel lines correspond to constant fractions
of the dense gas (n(H$_2$)$\geq 10^4$\cc) and their slopes are those found by
Lada, Lombardi \& Alves [16]. If one takes into account that the CO mass measurements
reflect an average on relatively large spatial scales in extragalactic systems ($\sim$ 1 kpc),
and corrects for this effect, such that the true mass involved in the star formation process
(that of the dense gas) is included in the SFR- dense gas mass diagram, then local galactic
molecular clouds and extragalactic system share the same relationship. According to
Lada et al. [17], "{\it there is a fundamental empirical scaling relation that directly connects
the local star forming process with that operating globally within galaxies}".

\begin{figure}
\sidecaption[t]
\includegraphics[scale=0.85]{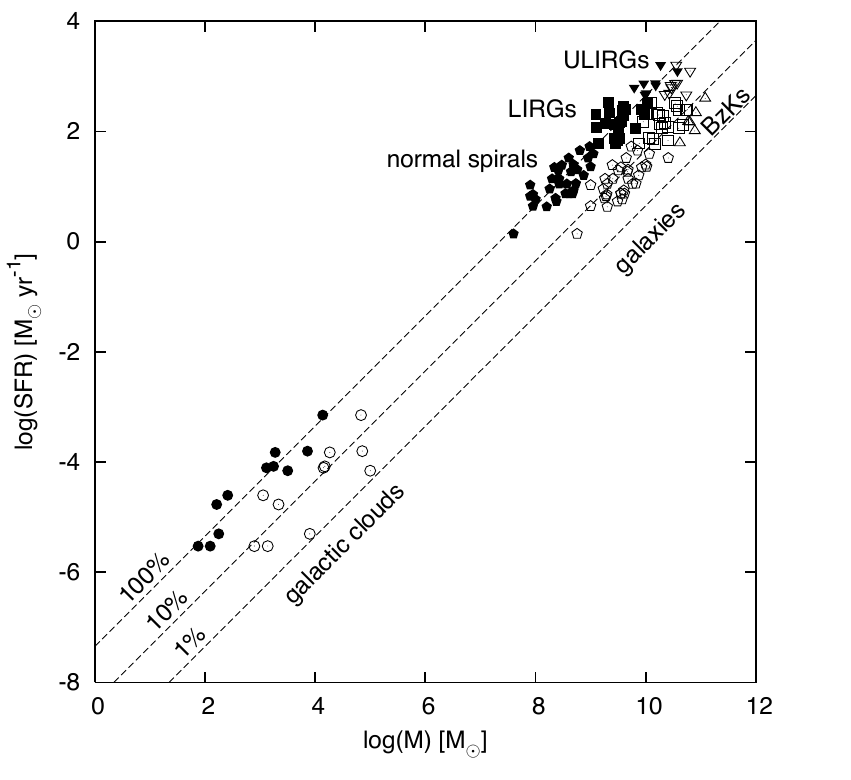}
%
%
\caption{The SFR-molecular cloud mass diagram according to
Lada et al. [17]). Open symbols correspond to "mean" molecular gas 
masses, while the solid ones to those obtained for a dense gas either
using extinction or HCN measurements (see text).}
\label{fig:6}       
\end{figure}

\section{Synthetic Modeling of the Milky Way}
\label{sec:4}

When studying extragalactic systems, either nearby or at a high redshift, a common technique
to interpret the observations is to compare them with synthetic models of a galaxy. In a model
one can modify the Initial Mass Function and rate of star formation, plus add more information
like a stellar spectra library to follow the spectral evolution of a system
 (see e.g. Bruzual \& Charlot [4, 5]). Recently this technique has been applied by 
 Robitaille \& Whitney (2010) [24] to estimate the star formation rate of the Milky Way, based on
 a comparison with the mid-IR data obtained by GLIMPSE plus MIPSGAL (Churchwell et al. [8];
 Carey et al.  [6]), and as an extension to the thorough modeling carried out 
 by Robitaille et al [22]  of the spectral energy distribution  (SED) of protostellar objects. 
 For the Milky Way, Robitaille \& Whitney  [24] looked at the 3D 
 distribution of star formation within the disk in such way to take into account not only their YSO
 theoretical SED models, but also including the limitations in sensitivity by the IRAC instrument
 as well as the effects of dust extinction. Although the technique is quite powerful, it does depend
 on the assumptions of the model. The prescription of Robitaille \& Whitney [22]  goes as follows:
 distribute the YSOs in random positions, use random age and mass given by the Kroupa [15]
 Initial Mass function (within 0.1 to 50\msol), control the upper and lower stellar age, include a
 reasonable spatial distribution of dust, use the synthetic SEDs to estimate the
 intrinsic magnitudes and IRAC colors, select only those YSOs that fall within the survey area 
 and fulfill the criteria of color and brightness defined by Robitaiille et al. [23]; and finally,
 adjust the SFR to match the observations. In practice, only sources that are younger than 2Myr
 are used to really match the observations.

Figures 7 and 8 show the results for one of these models, the mass distribution function
of the synthetic YSOs and their corresponding spatial distribution in the the Milky Way, 
respectively. The red histogram of the mass distribution (Fig. 7) corresponds to what is actually 
observed, and this "represents to less than 0.5\% of all the YSOs in the Galaxy" [24],
 i.e. the SFR is obtained from a very small fraction of objects. Also these YSOs
are systematically more massive with a median value of 10-15\msol, but within a 3 to 20\msol
range. The spatial distribution (Fig. 8) is color coded according to extinction along the 
line-of-sight from us, and shows that most of the sources that are counted  [23]
are within 10-15 kpc from us.

\begin{figure}[t]
\sidecaption[t]
\includegraphics[scale=0.85]{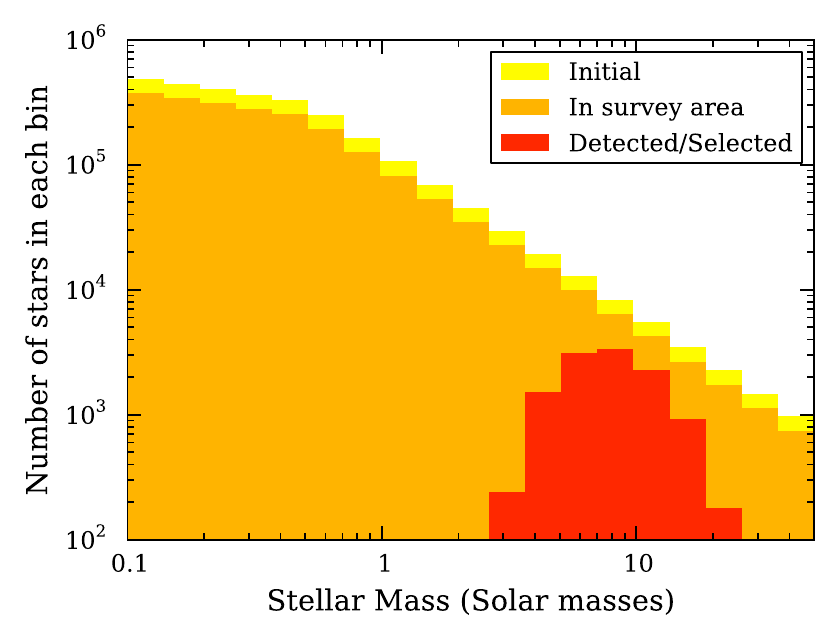}
\caption{The mass distribution function of the synthetic YSOs for
a given synthesis model. The yellow and orange histograms show
the initial and surveyed distributions, respectively; while the red one only those
YSOs detected by GLIMPSE and MIPSGAL  surveys (Robitaille \& Whitney [24]).
}
\label{fig:7}       
\end{figure}

\begin{figure}[t]
\sidecaption[t]
\includegraphics[scale=0.80]{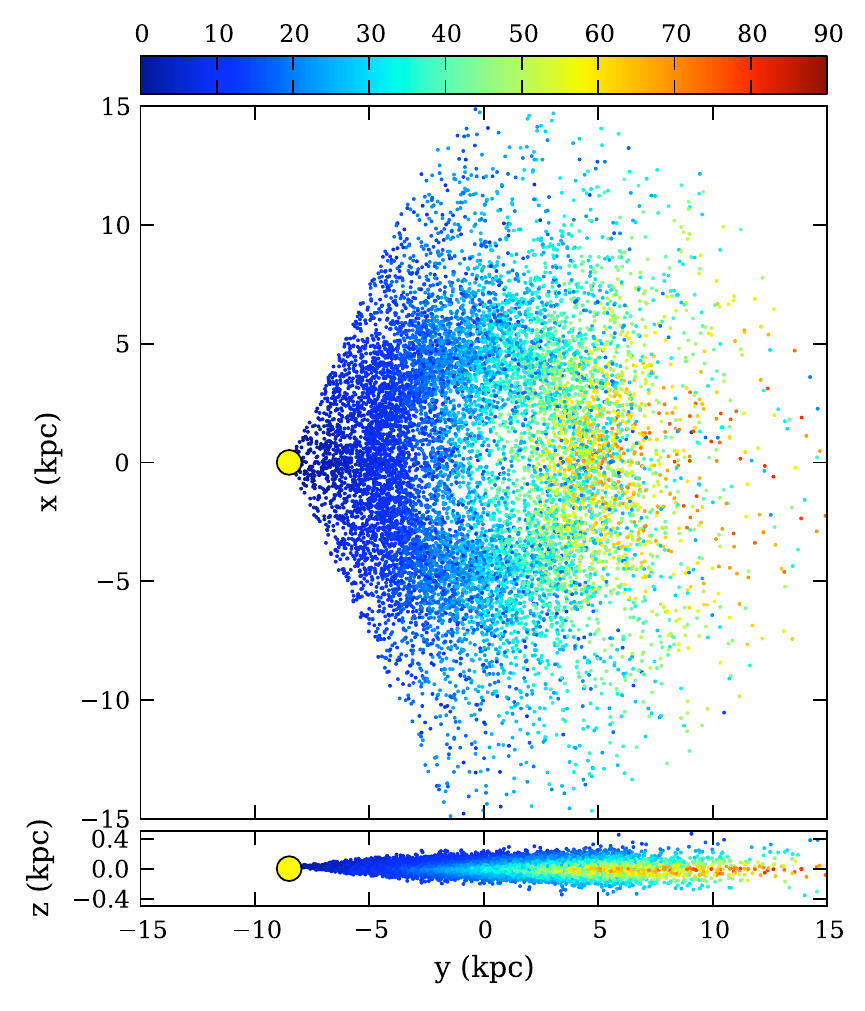}
\caption{The synthetic 3-dimensional spatial distribution of the YSOs
as detected by GLIMPSE \& MIPSGAL. The points are color-coded according
to the extinction values along-the-line of sight from us (see Robitaille
\& Whitney  [24] for details).
}
\label{fig:8}       
\end{figure}

 After several "realizations" of the synthetic models, and taking into account  that the 
contamination by AGB stars in the GLIMPSE \& MIPSGAL sample ranges from 30\% to 50\%,
the SFR in the Galaxy is estimated to be between 0.68 to 1.45\mdot  [24].

\section{Far Infrared Star Formation Rate Estimate}
\label{sec:5}

 The star formation rate estimated using mid-IR  observations, like that using the synthetic 
 models, does not take into account deeply embedded protostellar objects which are detected
 a wavelengths longer than 24\mum, and this should not be a problem for the far-IR
 observations. Recently Veneziani et al. [28] have used data from the Herschel 
 Space telescope of the HiGAL Key Project (Molinari et al. [20]) to estimate the rate of 
 star formation in the Milky Way. The HiGAL KP surveyed the inner Galaxy 
 (l = $\pm$ 60\arcdeg, b = $\pm$ 1\arcdeg) using the PACS and SPIRE instruments in parallel 
 mode, covering five wavelengths: 70,160, 250, 350 and 500\mum. At 70\mum~the angular 
 resolution is 6\arcsec, while at 500\mum~is $\sim$ 36\arcsec, and therefore, one of the biggest 
 challenges of this type of survey is to correctly bandmerge the flux densities of the compact 
 sources between short and longer wavelengths (Molinari et al. [21]). 
 This is an issue because it is quite possible that many massive protostars deeply embedded are
  surrounded by low mass protostars when forming, and the long wavelength observations cannot 
  resolve the multiple components (see e.g. [19]).
 
 Veneziani et al. used the data from the so called Science Demonstration Phase (SDP),
 taken at two latitudes, l=30\arcdeg~and l=59\arcdeg, where the conditions of star formation
 are quite different. The l=30\arcdeg~region is very active, including the well known W43 complex
 in its 2\arcdeg~field-of-view, while the l=59\arcdeg~field looks into the Vulpecula region, tangent
 to a spiral arm [3]. For both of these fields there are very good estimates of the
 distances [25], so mass envelope and bolometric luminosities for the compact
protostellar sources can be measured and be placed in the L$_{BOL}$ vs M$_{ENV}$ diagram
 (see e.g. [9, 19]). This diagram allowed Veneziani et al. [28] to
 follow the protostellar objects through their evolutionary tracks to their final zero
 age main sequence mass and count how many young stars per unit mass per unit time were
 found in the SDP fields. Preliminary results for the l=30\arcdeg~field found 690 sources, 
 with $\sim 323$ being likely to be protostars, within a range of mass envelope of 80 to
 2000\msol and a median mass of 540\msol.
These objects have a median evolutionary time of $\sim 2.5\times 10^5$ yr and a median final
main sequence mass of $\sim 15$\msol. Using Lada et al [16] relationship for the star 
formation rate using the median mass and evolutionary time, one gets
SFR = N$_{YSOs}$~M$_{final}$/t$_{evol}$ or ${323 \times 15\over 2.5\times 10^5}$ = 0.02\msol/yr. 
This rate is approximately 20 times that of Orion A and B [16], perhaps too high to
be representative of the entire Milky Way. This rate suggests that the simple method of counting
protostellar objects is more complicated when dealing with high mass protostars, and that the 
number of YSOs identified plus their median final mass and lifetime under the "accretion" 
formation scenario can be a bit uncertain.

\begin{figure}[t]
\sidecaption[t]
\includegraphics[scale=.28]{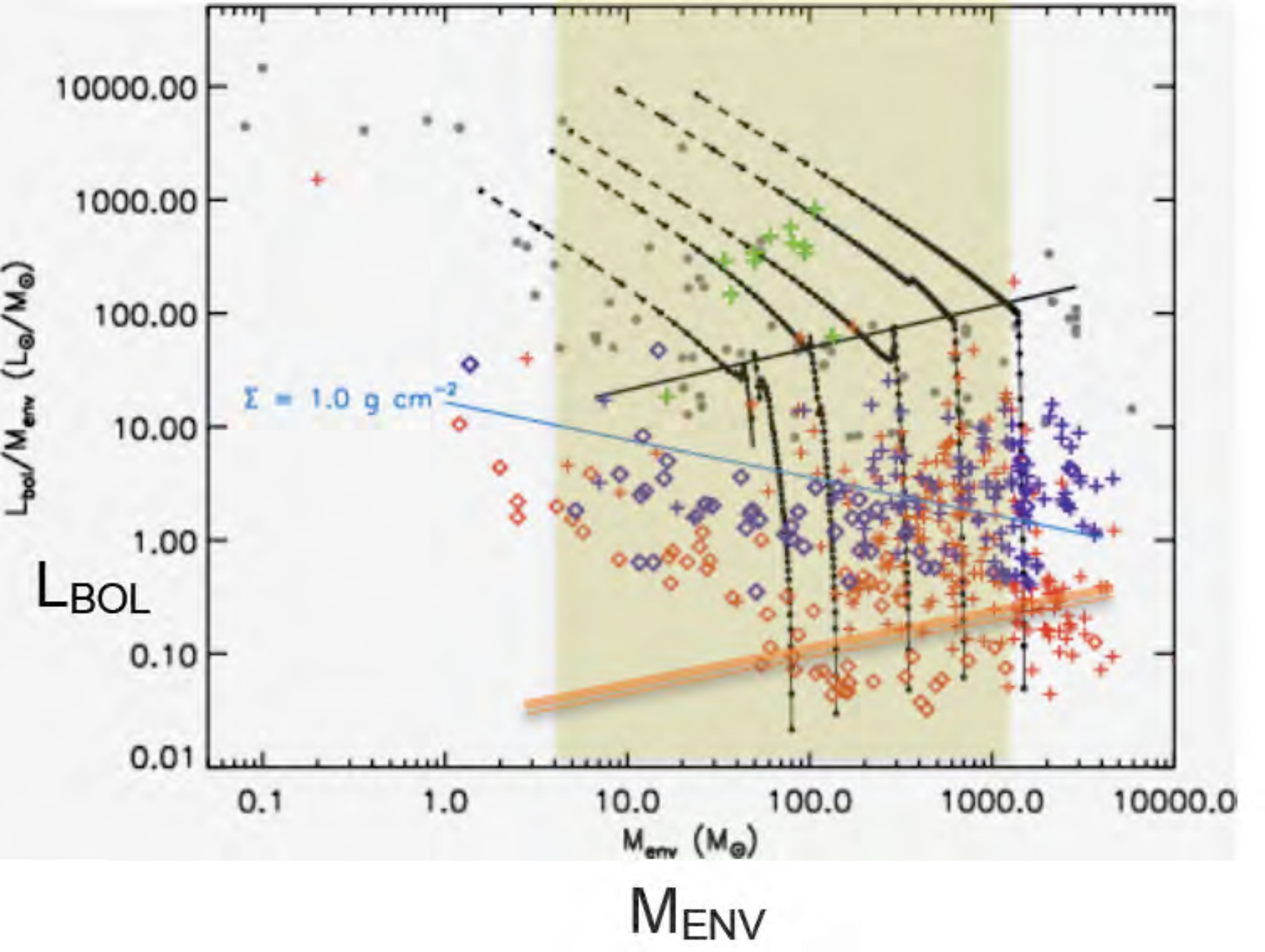}
%
%
\caption{YSOs found in the l=30\arcdeg~SDP field place in the massive star
formation bolometric luminosity (L$_{BOL}$) vs. mass envelope (M$_{ENV}$)
diagram (after Molinari et al. [19], Elia et al. [ 9]). The shaded region marks the range of
envelope masses considered by Veneziani et al. [28] in their SFR estimate.}
\label{fig:9}       
\end{figure}

Given the relatively high luminosity at 70\mum~of the l=30\arcdeg~SDP field,
and the fact that the selected YSOs were those found in the densest regions,
Veneziani et al. [28] considered a second approach to estimate the SFR, by
using an extrapolation of the FIR extragalactic indicator developed by Li et al. [18].
 The Li et al. 70\mum~SFR indicator is based on a sample of 40 SINGs galaxies
 and uses 'sub-galactic' regions with sizes between 0.05 and 2 kpc, and in this sense
 closer to the scales that are sampled in the Milky Way. The indicator is calibrated for
 a range of 70\mum~luminosities of  $5 \times 10^{40} \leq L(70) \leq 5\times 10^{43}$ \lum
  and given by, SFR(\mdot) = L(70)/$1.067\times 10^{43}$\lum.
  For the 70\mum~luminosity of the l=30\arcdeg~field, this corresponds to a SFR of
  $3.8\pm0.7\times 10^{-4}$\mdot.
  
  If one assumes that  HiGal l=30\arcdeg~tile is representative of the entire Milky Way,
 then is possible to have a rough estimate of the SFR rate in the Galaxy, by weighting
 the volume of each tile with respect to that of the Milky Way [28].
 Approximating the Galaxy as a simple disk with a scale heigh of 100pc  and a mean 
 radius of 15kpc, the total mean star formation rate is 2.1$\pm 0.4$ \mdot, i.e. within
 the range obtained by the synthetic modeling of the Milky Way.

\begin{acknowledgement}
The bulk of this contribution was written during two visits to Paris, I thank Francois Boulanger
and Ken Ganga for providing me with a working environment to complete it. I also thank the organizing
committee for their kind invitation.
\end{acknowledgement}
%
%
%
%

\end{document}